\documentclass[prd,aps,nofootinbib,onecolumn,notitlepage,longbibliography,unsortedaddress,superscriptaddress]{revtex4-1}

\usepackage[letterpaper,
  paperwidth=215.9mm,
  paperheight=279.4mm,
   top=25.4mm,
   headsep=10mm,
   bottom=25.4mm,
   footskip=18mm,
   inner=25.4mm,
   outer=25.4mm,
]{geometry}

\usepackage[T1]{fontenc}
\usepackage[utf8]{inputenc}
\usepackage{lmodern}

\usepackage{minibox}
\usepackage{boxedminipage}
\usepackage{needspace} 

\usepackage{multirow}
\usepackage{booktabs}
\usepackage{cellspace}

\usepackage{graphicx}
\usepackage{xcolor}
\definecolor{mylinkcolor}{rgb}{0.0,0.0,0.66}

\usepackage[plainpages=false,
pdftitle={On Seminal HEDP Research Opportunities Enabled by Colocating Multi-Petawatt Laser with High-Density Electron Beams},
pdfauthor={Sebastian Meuren, Phil H. Bucksbaum, Nathaniel J. Fisch, Frederico Fi{\'u}za, Siegfried Glenzer, Mark J. Hogan, Kenan Qu, David A. Reis, Glen White, Vitaly Yakimenko},
pdfcreator={LaTex 2e},
pdfkeywords={}
]{hyperref}
\hypersetup{colorlinks=true,linkcolor=mylinkcolor,urlcolor=mylinkcolor,citecolor=mylinkcolor,filecolor=mylinkcolor,breaklinks=true,linktocpage=true,linktoc=all}

\usepackage{bm}
\usepackage[intlimits]{amsmath}
\usepackage{amssymb}
\usepackage{upgreek}
\usepackage{slashed}
\usepackage{units}
\newcommand{\nfrac}[2]{{#1}/{#2}}

\newcommand{\average}[1]{\langle#1\rangle}
\newcommand{\eps}{\epsilon}
\newcommand{\cascadeparam}{\mathcal{C}_L}
\newcommand{\electronenergy}{\mathcal{E}}

\newcommand{\sdist}{\kern 0.20em}

\renewcommand{\eqref}[1]{Eq.\sdist(\ref{#1})}
\newcommand{\figref}[1]{Fig.\sdist\ref{#1}}
\newcommand{\tabref}[1]{Tab.\sdist\ref{#1}}

\makeatletter
\def\normalsize{%
   \fontsize{11}{13.2}\selectfont
   \abovedisplayskip 10\p@ \@plus2\p@ \@minus5\p@
   \belowdisplayskip \abovedisplayskip
   \abovedisplayshortskip  \abovedisplayskip
   \belowdisplayshortskip \abovedisplayskip
   \let\@listi\@listI
}%
\def\small{%
  \fontsize{10}{12}\selectfont
  \abovedisplayskip 8.5\p@ \@plus3\p@ \@minus4\p@
  \belowdisplayskip \abovedisplayskip
  \abovedisplayshortskip \z@ \@plus2\p@
  \belowdisplayshortskip 4\p@ \@plus2\p@ \@minus2\p@
  \def\@listi{%
    \leftmargin\leftmargini
    \topsep 4\p@ \@plus2\p@ \@minus2\p@
    \parsep 2\p@ \@plus\p@ \@minus\p@
    \itemsep \parsep
  }%
}%
\def\footnotesize{%
  \fontsize{10}{12}\selectfont
  \abovedisplayskip 6\p@ \@plus2\p@ \@minus4\p@
  \belowdisplayskip \abovedisplayskip
  \abovedisplayshortskip \z@ \@plus\p@
  \belowdisplayshortskip 3\p@ \@plus\p@ \@minus2\p@
  \def\@listi{%
    \leftmargin\leftmargini
    \topsep 3\p@ \@plus\p@ \@minus\p@
    \parsep 2\p@ \@plus\p@ \@minus\p@
    \itemsep \parsep
  }%
}%
\makeatother

\makeatletter                                                                 
 \def\footnoterule{\kern-6\p@    %
 \hrule \@width 2in \kern 5.7\p@}  %
\makeatother

\makeatletter 
\renewcommand{\fnum@figure}{\textbf{\textsf{Fig.\sdist\thefigure}}}
\renewcommand{\fnum@table}{\textbf{\textsf{Tab.\sdist\thefigure}}}
\makeatother

\begin{document}

\title{\textsf{On Seminal HEDP Research Opportunities Enabled by Colocating\\{}Multi-Petawatt Laser with High-Density Electron Beams\vspace*{5pt}}}

	\author{Sebastian Meuren}
	\thanks{\url{smeuren@stanford.edu}}
	\affiliation{\mbox{Stanford PULSE Institute, SLAC National Accelerator Laboratory, Menlo Park, CA 94025}}%
	\author{Phil H.\ Bucksbaum}%
	\affiliation{\mbox{Stanford PULSE Institute, SLAC National Accelerator Laboratory, Menlo Park, CA 94025}}%
	\author{Nathaniel J.\ Fisch}%
	\affiliation{\mbox{Department of Astrophysical Sciences, Princeton University, Princeton, NJ 08544}}%
	\author{Frederico Fi{\'u}za}%
	\affiliation{\mbox{SLAC National Accelerator Laboratory, Menlo Park, CA 94025}}%
	\author{Siegfried Glenzer}%
	\affiliation{\mbox{SLAC National Accelerator Laboratory, Menlo Park, CA 94025}}%
	\author{Mark J.\ Hogan}%
	\affiliation{\mbox{SLAC National Accelerator Laboratory, Menlo Park, CA 94025}}%
	\author{Kenan Qu}%
	\affiliation{\mbox{Department of Astrophysical Sciences, Princeton University, Princeton, NJ 08544}}%
	\author{David A.\ Reis}%
	\affiliation{\mbox{Stanford PULSE Institute, SLAC National Accelerator Laboratory, Menlo Park, CA 94025}}%
	\author{Glen White}%
	\affiliation{\mbox{SLAC National Accelerator Laboratory, Menlo Park, CA 94025}}%
	\author{Vitaly Yakimenko\vspace*{5pt}}%
	\affiliation{\mbox{SLAC National Accelerator Laboratory, Menlo Park, CA 94025}}%
\date{\today}

\enlargethispage{2\baselineskip}
\begin{abstract}
\vspace*{10pt}The scientific community is currently witnessing an expensive and worldwide race to achieve the highest possible light intensity. Within the next decade this effort is expected to reach nearly $\unitfrac[10^{24}] {W}{cm^2}$ in the lab frame by focusing of 100 PW, near-infrared lasers. A major driving force behind this effort is the possibility to study strong-field vacuum breakdown and an accompanying electron-positron pair plasma via a quantum electrodynamic (QED) cascade [Edwin Cartlidge, ``\textit{The light fantastic}'', Science 359, 382 (2018)]. Whereas Europe is focusing on all-optical 10 PW-class laser facilities (e.g., Apollon \& ELI), China is already planning on co-locating a 100 PW laser system with a 25\,keV superconducting XFEL and thus implicitly also a high-quality electron beam [Station of Extreme Light (SEL) at the Shanghai Superintense-Ultrafast Lasers Facility (SULF)]. This white paper elucidates the seminal scientific opportunities facilitated by colliding dense, multi-GeV electron beams with multi-PW optical laser pulses. Such a multi-beam facility would enable the experimental exploration of extreme HEDP environments by generating electron-positron pair plasmas with unprecedented densities and temperatures, where the interplay between strong-field quantum and collective plasma effects becomes decisive.
\end{abstract}

\maketitle

\textbf{\textsf{Relevance for astrophysics and cosmology.}} At the so-called QED critical or Schwinger field $E_{\text{cr}} = \nfrac{m^2c^3}{(\hbar e)} \approx \unitfrac[1.3 \times 10^{18}]{V}{m}$ the quantum vacuum becomes unstable with respect to electron-positron pair production \cite{di_piazza_extremely_2012,schwinger_gauge_1951}. Subsequently, the created charges are accelerated by the field and generate a ``short-circuit current'' which screens the field. This mechanism sets a fundamental limit for electric fields, e.g., in the vicinity of a space-time singularity like a black hole and thus significantly influences the dynamics of such objects \cite{parfrey_first-principles_2019,blandford_electromagnetic_1977}. 

Being stable with respect to pair production, magnetic fields may exceed the critical field strength $B_{\text{cr}} = m^2c^2/(\hbar e) \approx \unit[4.4 \times 10^9]{T}$, e.g., close to the surface of a neutron star \cite{olausen_mcgill_2014}. Even though the field itself is stable, energetic particles or photons can induce QED cascades in such strong fields \cite{cerutti_electrodynamics_2017}. During a QED cascade, i.e., a sequence of photon emission and pair production, the number of particles increases exponentially until an electron-positron pair plasma is formed. The resulting complex interplay between strong-field quantum and collective plasma effects is still only poorly understood \cite{chen_filling_2020,timokhin_maximum_2019,gueroult_determining_2019,melrose_pulsar_2016}. In particular, it is likely to qualitatively change the dynamic of magnetic reconnection and the resulting particle acceleration and radiation emission~\mbox{\cite{schoeffler_bright_2019,hakobyan_effects_2019,werner_particle_2019}}.   

Relativistic electron-positron pair plasmas have fundamentally different properties with respect to plasmas which consist of ionized atoms \cite{stenson_debye_2017,edwards_strongly_2016,ruffini_electronpositron_2010}. Due to their relevance for extreme astrophysical environments \cite{uzdensky_plasma_2014} and the primordial universe during the ``leptonic era'' \cite{kandus_primordial_2011,weinberg_cosmology_2008}, electron-positron pair plasmas in general and QED cascades in particular represent an exciting challenge for the next decade of plasma research (for more details see, e.g., the Astro2020 Science White Paper \textit{``Extreme Plasma Astrophysics''} \cite{uzdensky_extreme_2019} and the perspective article ``\textit{Relativistic Plasma Physics in Supercritical Field}'' \cite{zhang_relativistic_2020}).

\needspace{4\baselineskip}\textbf{\textsf{Relevance for light-matter interactions in extreme conditions.}} In a terrestrial laboratory electromagnetic fields are always much weaker than QED critical. A laser system, for example, would have to provide an intensity of $I_{\text{cr}} = c\eps_0 E^2_{\text{cr}}  \approx \unitfrac[4.6 \times 10^{29}]{W}{cm^2}$ in order to probe the QED critical field directly. Even though this intensity is clearly out of reach even in the mid-term future, $E_{\mathrm{cr}}$ is not inaccessible with existing technology. A relativistic charge experiences a rest-frame electric field $E^*\sim \gamma{}E$ which is enhanced by its Lorentz gamma factor $\gamma = \electronenergy/(mc^2)$. Therefore, strong-field quantum effects start to become decisive if $\chi = E^*/E_{\mathrm{cr}} \gtrsim 1$, i.e., as soon as a charge experiences the QED critical field in its rest frame~\cite{di_piazza_extremely_2012}. Therefore, $E_{\mathrm{cr}}$ can be probed by combining ultra-intense lasers with ulta-relativistic electron beams \cite{burke_positron_1997,bula_observation_1996}.

In the ``quantum regime'' $\chi \gtrsim 1$ a single photon emission induces a substantial recoil and thus disrupts the trajectory of the emitting charge. This type of ``non-adiabatic'' radiation friction force is different from the ``classical'' regime of radiation reaction ($\chi \ll 1$) in which, e.g., existing circular accelerators, synchrotrons, and FELs operate \cite{jackson_classical_1998}. Besides quantum modifications to synchrotron radiation electron-positron pair production becomes sizable if $\chi \gtrsim 1$ \cite{di_piazza_extremely_2012}. Intuitively speaking, typically emitted photons reduce the vacuum tunneling barrier to a level that allows the field to transform them into real electron-positron pairs \cite{dinu_trident_2018,meuren_semiclassical_2016,di_piazza_barrier_2009}. 

\begin{figure}[tb]
	\centering
    \includegraphics[width=1.0\textwidth]{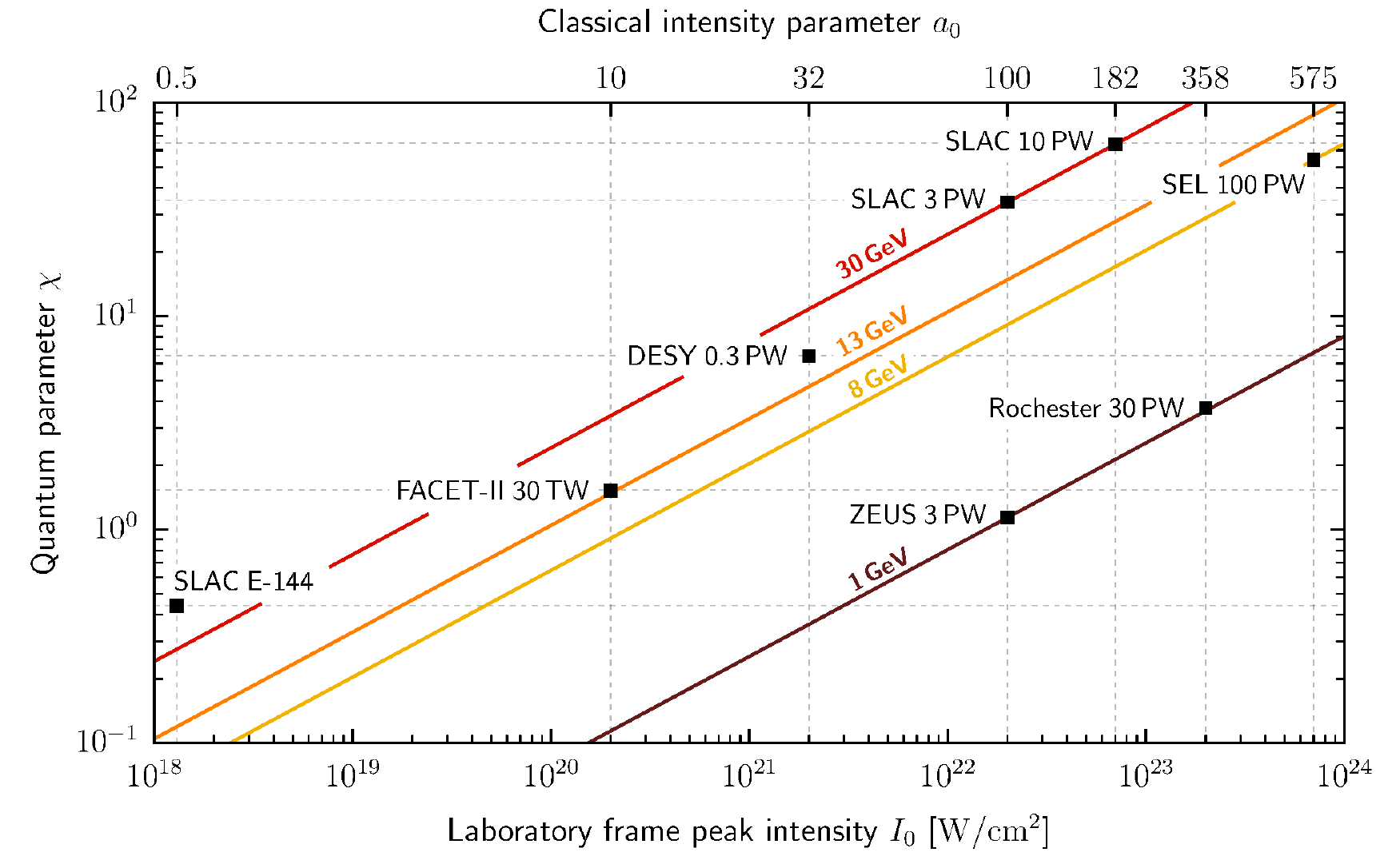}
	\caption{\label{fig:parameters}Quantum parameter $\chi$ as a function of the peak laser intensity $I_0$ and the electron beam energy $\electronenergy$. Different upcoming and proposed facilities are indicated (see \tabref{tab:facilities} for details). Evidently, the co-location of a multi-Petawatt laser system with the SLAC LINAC (30 GeV from combined FACET-II and LCLS-Cu) would provide a world-wide unique research opportunity.}
\end{figure}

Recent experiments at all-optical laser facilities have already seen the onset of quantum effects for synchrotron radiation in the regime $\chi\lesssim 0.1$ \cite{poder_evidence_2018,cole_experimental_2018}, and upcoming experiments, e.g., at DESY \cite{abramowicz_letter_2019} and FACET-II \cite{yakimenko_facet-ii_2019,sfqedatfacet_proposal_2018}, will study the regime $\chi\gtrsim 1$ in depth (see \figref{fig:parameters} and \tabref{tab:facilities}).

\begin{table}[p]
\heavyrulewidth=.08em
\lightrulewidth=.05em
\cmidrulewidth=.03em
\belowrulesep=.85ex
\belowbottomsep=0pt
\aboverulesep=.6ex
\abovetopsep=0pt
\cmidrulesep=\doublerulesep
\cmidrulekern=.5em
\defaultaddspace=3.5em
\tabcolsep=7pt
\extrarowheight=1pt
\newcommand{\tablefootintro}{\hspace*{-3.5pt}\fontsize{9}{10.8}\selectfont{}}
    \centering
    \begin{minipage}{\textwidth}\centering%
	\fontsize{9}{10.8}\selectfont{}
    \renewcommand{\thefootnote}{\textit{\alph{footnote}}}
	\begin{minipage}{\textwidth}\centering%
	\fontsize{10.5}{12.6}\selectfont{} %
    \begin{tabular}{lccc@{\hskip 20pt}c@{\hskip 20pt}c@{\hskip 20pt}cc}
        \toprule
        Facility & \minibox[c]{Laser\\{}Power} & \minibox[c]{Focal\\{}Area} & \minibox[c]{Peak Intensity\\{}(lab frame)} & \minibox[c]{Electron\\{}Energy} & $\chi$ & $a_0$ & $\cascadeparam\propto a_0^2$\\
        \midrule
        SLAC\,E-144\footnotemark[1] & ${\unit[0.4]{TW}}$ & $\unit[30]{\upmu{}m}^2$ & $\unitfrac[1.3\!\times\!10^{18}]{W}{cm^2}$ & \multirow{1}{*}{${\unit[46.6]{GeV}}$}  & $0.4$ & $0.5$ &  $\sim10^{-6}$\\
        \midrule
        FACET-II\footnotemark[2] & ${\unit[30]{TW}}$ & \multirow{3}{*}{$\unit[14]{\upmu{}m}^2$} & $\unitfrac[2\!\times\!10^{20}]{W}{cm^2}$ & \multirow{1}{*}{${\unit[13]{GeV}}$}  & ${1.5}$ & $10$ &  $6\!\times\!10^{-4}$\\
        DESY\footnotemark[3] & ${\unit[0.3]{PW}}$ & & $\unitfrac[2\!\times\!10^{21}]{W}{cm^2}$ & \multirow{1}{*}{${\unit[17.5]{GeV}}$} & $7$ & $32$ &  $6\!\times\!10^{-3}$\\
        ZEUS\footnotemark[4] & ${\unit[3]{PW}}$ &  & $\unitfrac[2\!\times\!10^{22}]{W}{cm^2}$ & \multirow{1}{*}{${\unit[1]{GeV}}$}  & ${1.2}$ & $100$ &  $6\!\times\!10^{-2}$\\
         \midrule
         & ${\unit[3]{PW}}$ & \multirow{4}{*}{$\unit[14]{\upmu{}m}^2$} & $\unitfrac[2\!\times{}\!10^{22}]{W}{cm^2}$ & \multirow{1}{*}{${\unit[30]{GeV}}$} & ${35}$ & $100$  &  $0.06$\\
         \multirow{1}{*}{SLAC\footnotemark[5]}  & $\boldsymbol{\unit[10]{PW}}$ &  & $\unitfrac[7\!\times{}\!10^{22}]{W}{cm^2}$ & \hspace*{-3.5pt}\multirow{1}{*}{$\boldsymbol{\unit[30]{GeV}}$\hspace*{-5pt}} & $\boldsymbol{65}$ & $182$ &  $0.2$\\
        Rochester\footnotemark[6] &${\unit[30]{PW}}$ &  & $\unitfrac[2\!\times{}\!10^{23}]{W}{cm^2}$ & \multirow{1}{*}{${\unit[1]{GeV}}$} & $4$ & $358$  &  $0.7$\\
        SEL\footnotemark[7] & $\boldsymbol{\unit[100]{PW}}$ &  & $\unitfrac[7\!\times{}\!10^{23}]{W}{cm^2}$ & \multirow{1}{*}{$\boldsymbol{\unit[8]{GeV}}$} & $\boldsymbol{55}$ & $575$  &  $2$\\
         \bottomrule
    \end{tabular}\vspace*{3pt}
	\end{minipage}
    \renewcommand{\thefootnote}{\textit{\alph{footnote}}}\setlength{\footnotesep}{9pt}
    \footnotetext[1]{\tablefootintro{}First experiment which approached $\chi\sim 1$ with electron-laser collisions in the 1990s \cite{burke_positron_1997,bula_observation_1996}.}
    \footnotetext[2]{\tablefootintro{}Approved SLAC E-320 experiment at FACET-II \cite{yakimenko_facet-ii_2019,sfqedatfacet_proposal_2018}.}
    \footnotetext[3]{\tablefootintro{}LUXE experiment proposed at DESY \cite{abramowicz_letter_2019}.}
    \footnotetext[4]{\tablefootintro{}ZEUS laser system approved at the University of Michigan \cite{ZEUS} if combined with a $\unit[1]{GeV}$ RF\,LINAC.}
    \footnotetext[5]{\tablefootintro{}A ${\unit[30]{GeV}}$ electron beam is achievable with existing infrastructure at SLAC by combining the FACET-II and the LCLS-Cu RF\,LINAC (time scale including new building: 2-3 years).}
    \footnotetext[6]{\tablefootintro{}Laser system proposed in Rochester \cite{bromage_technology_2019} if combined with a $\unit[1]{GeV}$ RF\,LINAC.}
    \footnotetext[7]{\tablefootintro{}Station of Extreme Light (SEL) approved at the Shanghai Superintense-Ultrafast Laser Facility (SULF) \cite{cartlidge_light_2018}. Note that the challenges of co-locating the high-intensity laser focus with the XFEL electron beam and achieving a $\unit[3]{\upmu{}m}$ focus at 100\,PW have not been resolved yet. }
    \end{minipage}
    \vspace*{2pt}\caption{\label{tab:facilities}\fontsize{10.5}{12.6}\selectfont{}\textbf{\textsf{Capabilities of different facilities}} (see also \figref{fig:parameters}).\ Access to the deep quantum regime ${\chi\gg 1}$ would facilitate the investigation of beam-driven QED cascades, high-density electron-positron pair plasmas, and light-matter interactions in a qualitatively different regime. Notably, the multiplicity of a QED cascade scales proportional to $\chi$, i.e., $\chi \sim 10$ implies that the density of the produced pair plasma exceeds the density of the driving beam by a factor of $\sim10$ \cite{qu_observing_2020}. %
    An alternative approach, which provides access to complementary physics, is to study self-sustained QED cascades with laser-laser collisions. The corresponding threshold is ${\cascadeparam\gtrsim 1}$  (i.e., $a_0\gtrsim 400$ for optical laser systems). Note that we assume the same focal area $A = \pi w_0^2/2 \approx \unit[14]{\upmu{}m^2}$ for all upcoming facilities ($w_0 = \unit[3]{\upmu{}m}$). Empirically, however, it is more challenging to obtain a small spot size with an extreme-power laser system, which usually has a low repetition rate (state-of-the-art spot size: $w_0 = \unit[2.4]{\upmu{}m}$ achieved with a $\unit[100]{TW}$-class laser system \cite{yan_high-order_2017}). For an ideal Gaussian laser pulse with cycle-averaged peak power $P$ the cycle-averaged peak intensity $I_0$ is given by $I_0 = P/A$ \cite{cros_laser-driven_2016}. Assuming linear laser polarization and a head-on collision with the electron beam the sub-cycle peak values of $\chi$, $a_0$, and $\cascadeparam$ are given by $\chi = \unit[0.057]{GeV}^{-1} \electronenergy  \sqrt{2I_0/(\unit[10^{20}]{Wcm^{-2}})}$, $a_0 = \unit[0.60]{\upmu{}m}^{-1} \lambda_L  \sqrt{2I_0/(\unit[10^{18}]{Wcm^{-2}})}$, and $\cascadeparam = 2a^2_0 \, [\hbar\omega/(mc^2)]$, respectively. Here, $\electronenergy$ is the electron energy and we assume a laser wavelength of $\lambda_L = \unit[800]{nm}$ except for SLAC\,E-144 ($\lambda_L = \unit[527]{nm}$) \cite{burke_positron_1997} and Rochester ($\lambda_L = \unit[910]{nm}$)~\cite{bromage_technology_2019}.
    \newline{}%
    Note that it is preferable to increase the electron energy over increasing the laser intensity if large values of the quantum parameter $\chi$ should be obtained (linear vs.\ square-root scaling): whereas the FACET-II $\unit[13]{GeV}$ electron beam only requires a $\unit[30]{TW}$ laser system to reach the Zettawatt power scale in the electron rest frame, $\unit[3]{PW}$ are necessary for $\unit[1]{GeV}$ electrons at the ZEUS laser in Michigan. Similarly, the combination of $\unit[10]{PW}$ and $\unit[30]{GeV}$ electrons at SLAC would achieve higher values for $\chi$ than the SEL $\unit[100]{PW}$ laser system under construction in China \cite{cartlidge_light_2018}. Here, we have assumed that this laser is combined with the $\unit[8]{GeV}$ XFEL electron beam available at the same facility \cite{huang_physical_2019}. According to existing plans, however, the 100\,PW laser can only interact with the x-rays, not with the electron beam itself. Note that LWFA electron beams with up to $\unit[8]{GeV}$ have been demonstrated \cite{gonsalves_petawatt_2019}. As explained in the main text, however, state-of-the-art RF\,LINAC beams achieve densities which are $\gtrsim 10^3$ higher. Only such extreme densities facilitate investigations of the interplay between strong-field quantum and collective plasma effects \cite{qu_observing_2020}.}
\end{table}

So far all theoretical predictions for the regime $\chi \gtrsim 1$ rely on numerical Monte Carlo codes~\cite{gonoskov_extended_2015}, which employ untested approximations \cite{blackburn_radiation_2019,di_piazza_implementing_2018}. The inclusion of polarization effects is recent \cite{li_ultrarelativistic_2019,del_sorbo_spin_2017,meuren_quantum_2011} and the influence of radiative corrections is completely missing. Electron-positron pair production is modeled only in the simplest way, even though pair production becomes a major part of the physics for $\chi\gtrsim 1$. Formation length effects in photon radiation and quantum coherence between produced pairs are still largely ignored, even though these effects potentially play an important role in the deep quantum regime. All in all, the process of quenching large electromagnetic fields by pair creation is only poorly understood. All of these issues need to be studied experimentally. We expect substantial back-and-forth discussions between experiment, theory, and simulation.

\vspace*{1\baselineskip}
\textbf{\textsf{Relevance for HEDP:\ beam-driven QED cascades and pair plasmas.}} Reaching the extreme quantum regime $\chi \gg 1$ would facilitate the production of electron-positron pair plasmas via beam-driven QED cascades. A laboratory experiment under such unprecedented extreme conditions would therefore enable detailed investigations of the interplay between strong-field quantum and collective plasma effects. Entering this qualitatively different ``quantum plasma regime'' requires the co-location of multi-PW laser systems with dense multi-GeV electron beams (see \figref{fig:parameters} and \tabref{tab:facilities}). In a beam-driven QED cascade, the driver initiates a sequence of photon emission and pair production steps that leads to an exponential increase in the number of particles. This process continues until pair production and hard gamma radiation lowers the effective fields to restore $\chi \lesssim 1$ \cite{qu_observing_2020}. Correspondingly, the higher the achievable initial value of $\chi$, the larger the multiplicity in the electron-positron cascade that could be observed in the experiment. In this aspect, the physics of beam-driven QED cascades is very similar to the situation encountered in the magnetosphere of magnetars \cite{chen_filling_2020,timokhin_maximum_2019,gueroult_determining_2019,melrose_pulsar_2016}. 

In principle, it would be possible to initiate beam-driven QED cascades with electron beams obtained from laser wakefield acceleration (LWFA) at all-optical laser facilities. However, the parameters of existing LWFA beams are currently not sufficient for achieving the extreme electron beam densities required to study the interplay between strong-field quantum and collective plasma effects. To illustrate this, we compare the parameters of the energy-record $\unit[8]{GeV}$ LWFA electron beam reported in \cite{gonsalves_petawatt_2019} with those provided by the FACET-II RF\,LINAC \cite{yakimenko_facet-ii_2019}. The smallest achievable beam radius $\sigma = \sqrt{\eps\beta}$  is determined by the beta function $\beta$ of the focusing system and the transverse geometric emittance $\eps$ of the beam itself. This implies that for comparable final focus systems and longitudinal emittances the figure of merit is given by $Q/\eps$, as this quantity determines the highest achievable beam density%
\footnote{Note, however, that with increasing pointing and beam energy jitter it becomes more challenging to achieve a small beta function with realistic final focus systems. Therefore, the design of a small beta function focusing system, which can tolerate both energy and pointing jitter of typical LWFA beams, represents an additional challenge of the all-optical approach.}. %
Here $Q$ denotes the total beam charge. Considering only LWFA accelerated electrons which are in the energy range of $\unit[7.5-8]{GeV}$ in \cite{gonsalves_petawatt_2019}, we find $Q=\unit[5]{pC}$ ($\sim 
1\%$ of the total recorded charge) and a normalized emittance of%
\footnote{The geometric transverse emittance is defined as $\eps = \sqrt{\average{x^2}\average{x'^2}-\average{xx'}^2}$, where $x$ is the transverse position and $x'=\theta$ the divergence angle of the beam particles ($\theta=p_x/p_z$, where $p_x$ is the transverse and $p_z$ the momentum in propagation direction). In the following, we assume ``matched conditions'', i.e., that the emittance of the beam neither grows during acceleration nor in the down-ramp of the plasma \cite{ariniello_transverse_2019,corde_femtosecond_2013,mehrling_transverse_2012}. This implies that $\average{xx'}^2=0$ and $\theta^2/\eps = (\omega_p/c) (1/\sqrt{2\gamma})$, i.e., $\eps = \theta^2 (c/\omega_p) \sqrt{2\gamma}$. For a relativistic gamma factor of $\gamma \approx 1.6 \times 10^4$, experimental measured divergence angle of $\theta \approx \unit[0.2]{mrad}$, and a plasma frequency of $\hbar\omega_p \approx \unit[0.02]{eV}$ we find a normalized transverse emittance of $\gamma\eps \approx \unit[1]{\upmu{}m}$ ($\hbar\omega_p \approx \unit[1.174]{eV} \sqrt{n/\unit[10^{21}]{cm^{-3}}}$, the reported plasma density was $n \approx \unit[3\times 10^{17}]{cm^{-3}}$, and $\hbar c \approx \unit[0.1973]{eV\,\upmu{}m}$)~\cite{gonsalves_petawatt_2019}.}%
~$\gamma\eps\gtrsim\unit[1]{\upmu{}m}$. A state-of-the-art RF\,LINAC provides $Q\approx\unit[2]{nC}$ and $\gamma\eps\approx\unit[3]{\upmu{}m}$ at $\unit[13]{GeV}$ \cite{yakimenko_facet-ii_2019} with comparable longitudinal emittances and can therefore deliver at least $\sim 10^3$ times higher beam densities for $\gtrsim\unit[10]{GeV}$ beam energies.

\enlargethispage{-2\baselineskip}
\vspace*{1\baselineskip}
\textbf{\textsf{Relevance for laser-matter and laser-laser interactions.}} QED cascades and electron-positron pair plasmas could also be produced in laser-matter \cite{ridgers_dense_2012} or seeded laser-laser collisions \cite{tamburini_laser-pulse-shape_2017,grismayer_seeded_2017,bell_possibility_2008}. A convenient gauge and Lorentz invariant measure of laser intensity is provided by the classical intensity parameter $a_0 = \nfrac{e\!E}{(m\omega c)}$. Here, $E$ and $\hbar\omega$ denote the peak field strength of the laser and the characteristic energy of its photons, respectively. Plasma electrons typically exhibit a Lorentz gamma factor of $\gamma \sim a_0$ \cite{kruer_physics_2003}. As 
$\chi \approx 2\gamma[\hbar\omega/(mc^2)]\, a_0$, it is convenient to introduce the laser-laser cascade parameter $\cascadeparam = 2a^2_0 \, [\hbar\omega/(mc^2)]$ \cite{grismayer_seeded_2017}. If $\cascadeparam\gtrsim 1$ most plasma electrons experience the QED critical field in their rest frame, which implies that strong-field quantum effects start to play a decisive role in laser-plasma interactions even if the plasma is at rest in the laboratory frame%
\footnote{Note that $\chi = \cascadeparam$ for a plasma electron with gamma factor $\gamma = a_0$ which is counter-propagating with respect to the laser.}.

Note that reaching $\cascadeparam \gtrsim 1$ requires $a_0 \gtrsim 400$ [$2a_0^2 \gtrsim mc^2/(\hbar\omega)$], i.e., intensities $I \gtrsim \unitfrac[10^{24}]{W}{cm^2}$ (assuming an optical laser, i.e., $\hbar\omega \sim \unit[1]{eV}$) \cite{tamburini_laser-pulse-shape_2017,grismayer_seeded_2017,bell_possibility_2008}. With a $\unit[10]{PW}$ laser system this intensity is barely reachable (required focal area: $\unit[1]{\upmu{}m^2}$), but it is assumed that at least the onset of QED cascades will be accessible in laser-laser interactions at $\unit[10]{PW}$ laser systems like Apollon and ELI \cite{danson_petawatt_2019,eliwhitebook}.

An extensive investigation of QED cascades with laser-laser collisions, however, requires a $\unit[100]{PW}$-class laser system like the one proposed in Rochester \cite{bromage_technology_2019} or the one approved for construction in China \cite{cartlidge_light_2018} (see \tabref{tab:facilities} and \figref{fig:parameters} for details). Note that it is highly non-trivial to focus 100 PW laser power, necessarily distributed among multiple beams, to a focal area of $\unit[10]{\upmu{}m^2}$ in order to reach intensities $I \unitfrac[\gtrsim 10^{24}]{W}{cm^2}$. Solving this challenge will require substantial R\&D beyond the current state-of-the-art \cite{barty_nexawatt_2016}.

\vspace*{1\baselineskip}
\textbf{\textsf{Relevance for future linear collider.}} A linear lepton collider for the energy frontier of particle physics has to provide both very high center-of-mass energies and very high luminosities \cite{shiltsev_modern_2019} (CLIC: up to $\unit[3]{TeV}$ and $\sim\unit[10^{34}]{cm^{-2}s^{-1}}$ \cite{clic_2018}). As a result, the space charge at the interaction point will be extreme and the quantum regime $\chi\gtrsim 1$ is entered. Correspondingly, beamstrahlung energy losses during the collision represent a severe design limitation \cite{yokoya_beam-beam_1992}. CLIC, for example, reaches $\chi \sim 10$ at $\unit[3]{TeV}$, even though the design tries to minimize the value of the quantum parameter, e.g., by employing flat and long bunches, in order to reduce beamstrahlung as much as possible~\cite{esberg_strong_2014}. 

Recently, it has been suggested in \cite{yakimenko_prospect_2019,yakimenko_exhilptalk_2017} that beamstrahlung can also be mitigated by employing short and round bunches. This maximizes the quantum parameter ($\chi \gtrsim 10^3$ already at $\sim\unit[100]{GeV}$ with $\unit[100]{nm}$ bunches), but due to the short interaction time only very few leptons emit photons and thus lose energy \cite{harvey_quantum_2017,meuren_quantum_2011}. This approach implies a drastic reduction of the required beam/wall power, which significantly decreases the costs and implies scalability to the ``discovery regime'' ($\unit[10]{TeV}$, $\unit[10^{36}]{cm^{-2}s^{-1}}$). 

Note that the photon yield of beam-beam collisions is controllable by tuning the interaction time. In the extreme quantum limit $\chi\gg 1$, the beamstrahlung photon spectrum is expected to peak at the highest energies, energies comparable to the original electron energies. Then, beamstrahlung generated in electron-electron collisions could be used to realize a laserless gamma-gamma collider \cite{tamburini_efficient_2019,telnov_problems_1990,blankenbecler_quantum_1988}. On the other hand, allowing a sufficiently long interaction length would induce a full-featured QED cascade, which would result in a dense electron-positron pair plasma \cite{yakimenko_prospect_2019,del_gaudio_bright_2019}. With this tuning, the collider could also be used as an exciting HEDP platform to study pair plasmas under extreme conditions.

\vspace*{1\baselineskip}
\textbf{\textsf{Relevance for the strong-field frontier.}} In extremely strong electromagnetic background fields ($\chi \ggg 1$) leading-order radiative corrections scale as $\alpha\chi^{2/3}$ and a full breakdown of perturbation theory has been conjectured for $\alpha\chi^{2/3} \gtrsim 1$ ($\chi\gtrsim 10^3$) \cite{fedotov_conjecture_2017,ritus_radiative_1972}. The collider suggested in \cite{yakimenko_prospect_2019,yakimenko_exhilptalk_2017} aims at exploring this regime experimentally. This prospect has triggered a renewed interest in this old but so-far unsolved problem of quantum field theory in the presence of strong background fields (see, e.g.,  \cite{lavelle_renormalization_2019,ilderton_note_2019,podszus_high-energy_2019} and \cite{di_piazza_testing_2020,baumann_probing_2019,blackburn_reaching_2019}), which is of high relevance to beamstrahlung mitigation in linear collider using short and round bunches. 

The facility suggested here, that reaches $\chi\sim 100$ with laser-beam collisions, could both scrutinize state-of-the-art numerical codes for the description of matter and light in such extreme electromagnetic fields and substantiate the so-called Ritus-Narozhny conjecture by verifying scaling laws for radiative corrections.

\vspace*{1\baselineskip}
\begin{center}
\textbf{\textsf{Science Case: Beam-Driven QED Cascades and Pair Plasmas in Extreme Conditions}} 
\end{center}
\vspace*{0.2\baselineskip}

The experimental investigation of beam-driven QED cascades requires a quantum parameter of $\chi\gtrsim 10\!-\!100$ \cite{qu_observing_2020}. Due to the difference in scaling (linear vs. square root) it is preferable to increase the electron beam energy rather than the laser intensity for achieving large values of $\chi$ (see \figref{fig:parameters} and \tabref{tab:facilities}). Correspondingly, it is highly beneficial to co-locate high-power laser systems with high-energy particle beams.

Furthermore, it turns out that one needs extreme beam densities in order to study the interplay between strong-field quantum and collective plasma effects \cite{qu_observing_2020}. As pointed out above, state-of-the-art RF\,LINACs provide $\gtrsim 10^3$ higher beam densities if compared to state-of-the-art laser wakefield accelerators (LWFA). 

Therefore, we consider a high-density $\unit[30]{GeV}$ RF\,LINAC electron beam combined with a multi-PW laser system in the following. This energy scale could be reached, e.g., at SLAC, by combining the FACET-II with the LCLS-Cu RF\,LINAC \cite{yakimenko_facet-ii_2019}. For definiteness, we consider $P=\unit[3]{PW}$ ($\unit[10]{PW}$) laser pulses with a conservative spot size of $w_0 = \unit[3]{\upmu{}m}$, i.e., a focal area of $A = \pi w_0^2/2\approx\unit[14]{\upmu{}m^2}$. Thus, we only assume peak intensities $I_0=P/A$ up to $\unitfrac[2\!\times\!10^{22}]{W}{cm^2}$ ($\unitfrac[7\!\times\! 10^{22}]{W}{cm^2}$), i.e., $a_0 \approx 100$ ($a_0 \approx 180$). In combination with $\unit[30]{GeV}$ electrons $\chi \approx 35$ ($\chi \approx 65$) becomes accessible in head-on collisions (see \figref{fig:parameters} and \tabref{tab:facilities}).  

Note that the intensities considered here are more than one order of magnitude lower than the intensities required for studying QED cascades in laser-laser collisions ($\cascadeparam \gtrsim 1$, i.e., $I\gtrsim\unitfrac[10^{24}]{W}{cm^2}$) \cite{tamburini_laser-pulse-shape_2017,grismayer_seeded_2017}. Correspondingly, beam-driven QED cascades are much more accessible. Note however, that both approaches are complementary and provide access to different physics.

The typical radiation length of $\unit[30]{GeV}$ electrons is $\sim\gamma\lambdabar_c/(\alpha\chi^{2/3}) \lesssim \unit{\upmu{}m}$ (for $\chi \sim 10\!-\!100$)~\cite{yakimenko_prospect_2019}, where  $\lambdabar_c=\hbar/(mc) \approx \unit[3.9 \times 10^{-13}]{m}$ denotes the reduced Compton wavelength. Therefore, a head-on collisions between a $\unit[30]{GeV}$ electron beam and a $\gtrsim \unit[30]{fs} \approx \unit[10]{\upmu{}m}/c$ laser pulse, which is much longer than the $\unit[30]{GeV}$ radiation length, will induce a QED cascade. Since pair production does not stop until the effective fields are reduced to $\chi \lesssim 1$, the multiplicity of the cascade will be $\sim \chi$ \cite{qu_observing_2020}. This implies that already with a $\unit[3]{PW}$ laser the electron beam will produce an electron-positron pair plasma which has a density that is more than one order of magnitude higher than the original beam density. 

Collective plasma effects will start to influence the high-intensity optical laser itself if the effective plasma frequency $\omega_p$ becomes comparable to the laser frequency $\omega$, i.e., if $\sqrt{n/(\gamma\, 10^{21}\unit{cm^{-3}})} \sim 1$. Assuming a round electron beam with $\unit{nC}$ charge and $\unit{\upmu{}m}$ radius, final pair plasma densities $n \gtrsim \unit[10^{22}]{cm^{-3}}$ are achievable with a $\unit[3\!-\!10]{PW}$ laser system \cite{qu_observing_2020}. Due to first quantum and later also classical synchrotron radiation the final gamma factor $\gamma_f$ of the produced pair plasma is at least a factor $\sim 10\chi$ smaller than the gamma factor $\gamma_i$ of the initial beam \cite{qu_observing_2020}. Correspondingly, we obtain $\gamma_f \lesssim 10^2$ and $\omega_p \sim \omega$ for the parameters discussed here. 

Reaching the seminal milestone $\omega_p \sim \omega$ thus requires both $\chi\sim 10\!-\!100$ and electron beam densities $n \gtrsim \unit[10^{20}]{cm^{-3}}$. A facility with these parameters could observe the interplay between strong-field quantum effects (``non-adiabatic'' radiation reaction, pair production) and collective plasma effects (e.g., Weibel and two-stream instability; frequency upconversion in the driving laser pulse) for the first time \cite{qu_observing_2020}. 

Notably, the radiation pressure of the laser can stop and even reflect the pair plasma as soon as the ``reflection condition'' $\gamma \approx a_0/2$ is fulfilled \cite{qu_observing_2020,li_attosecond_2015}. For a $\unit[30]{GeV}$ electron beam this becomes possible between $\unit[3-10]{PW}$.

Inducing a QED cascade with an electron beam has the decisive advantage that one gains a high degree of control. By changing the parameters of the electron beam (energy, density, etc.) the dynamic of the QED cascade is steerable, which facilitates the scrutinization of numerical simulations and analytical scaling laws. Note that QED cascades in laser-laser collisions need to be seeded, and the details of the non-trivial seeding process have a significant impact on the cascade itself \cite{tamburini_laser-pulse-shape_2017}.

By colliding laser and electron beam at 90 degrees, which lowers the quantum parameter only by a factor of two, the interaction time could be substantially reduced \cite{blackburn_reaching_2019}. Therefore, already a $\gtrsim\unit[3]{PW}$ laser system would be sufficient to reach $\chi \gtrsim 10$ in this geometry. The short interaction time prevents a full-featured QED cascade and thus facilitates a measurement of photon emission and pair production in the deep quantum regime.

\vspace*{1\baselineskip}
\textbf{\textsf{In summary}}, the co-location of a multi-PW laser system with a high-density and high-energy electron beam enables seminal research opportunities for plasma physics in general and HEDP in particular. This has been highlighted, for example, in the {Executive Summary and Recommendations} section of the 2020 {Brightest Light Initiative Workshop Report}, which lists co-location as one of seven high-level recommendations: ``\textit{Collocating high-intensity lasers with other scientific infrastructure, such as facilities with relativistic particle beams or other energetic drivers, will enable forefront science in areas such as non-linear quantum electrodynamics, nuclear, plasma and high energy density physics, and astrophysics}'' \cite{bli_report_2020}.

\vspace*{0.5\baselineskip}
A facility which combines high-density multi-GeV electron beams with multi-PW optical laser would have decisive qualitative advantages in comparison with existing and upcoming all-optical laser facilities. Notably, such a facility is currently not planned anywhere in the world and would therefore provide a unique opportunity for the U.S.\ to re-claim leadership at the high-intensity laser frontier.
 
\vspace*{12\baselineskip}
\textbf{\textsf{Acknowledgments}}\vspace*{0.75\baselineskip}\\
\noindent{}The authors  thank Roger Blandford, Sébastien Corde, Elias Gerstmayr, and Michael Peskin for valuable discussions and suggestions. This work was supported by the U.S.\ Department of Energy under contract number DE-AC02-76SF00515. PHB and DAR were supported by the U.S.\ Department of Energy, Office of Science, Office of Fusion Energy Sciences under award DE-SC0020076. NJF and KQ were supported by NNSA Grant No.\ DE-NA0002948, and AFOSR Grant No.\ FA9550-15-1-0391. FF was supported by the U.S.\ DOE Early Career Research Program under FWP 100331. SG was supported by U.S.\ DOE Office of Science, Fusion Energy Sciences under FWP 100182.
\clearpage

\textbf{\textsf{References}}\vspace*{-0.2\baselineskip}\\

\end{document}